# Field emission from atomically thin edges of reduced graphene oxide

Hisato Yamaguchi[1]*[+], Katsuhisa Murakami[2]*[|], Goki Eda[3], Takeshi Fujita[4], Pengfei Guan[4], Weichao Wang[5], Cheng Gong[5], Julien Boisse[1||], Steve Miller[1], Muge Acik[5], Kyeongjae Cho[5], Yves J. Chabal[5], Mingwei Chen[4], Fujio Wakaya[2], Mikio Takai[2] and Manish Chhowalla[1+]

[1] Department of Materials Science and Engineering, Rutgers University, 607 Taylor Road, Piscataway, NJ 08854, U.S.A.

[2] Center for Quantum Science and Technology under Extreme Conditions, Osaka University, 1-3 Machikaneyama, Toyonaka, Osaka 560-8531, Japan

[3] Department of Materials, Imperial College London, Exhibition Road, London SW7 2AZ, UK.

[4] WPI Advanced Institute for Materials Research, Tohoku University, 2-1-1 Katahira, Aoba-ku, Sendai, Miyagi 980-8577, Japan

[5] Department of Materials Science and Engineering, University of Texas at Dallas, Richardson, Texas 75080, U.S.A.

*These authors contributed equally.

+Email: hisatoy@rci.rutgers.edu, manish1@rci.rutgers.edu

Present address:
[|] MATEIS, umr CNRS 5510, INSA de Lyon, 20 av. Albert Einstein, F-69621 Villeurbanne, Cedex, France
[||] Department of Electrical, Electronic and Communication Engineering, University of Erlangen-Nuremberg, Cauerstrasse 6, 91058 Erlangen, Germany

**ABSTRACT**

Point sources exhibit low threshold electron emission due to local field enhancement at the tip. The development and implementation of tip emitters have been hampered by the need to position them sufficiently apart to achieve field enhancement, limiting the number of emission sites and therefore the overall current. Here we report low threshold field (< 0.1V/μm) emission of multiple electron beams from atomically thin edges of reduced graphene oxide (rGO). Field emission microscopy (FEM) measurements show evidence for interference from emission sites that are separated by a few nanometers, suggesting that the emitted electron beams may be coherent. Based on our high-resolution transmission electron microscopy, infrared spectroscopy and simulation results, field emission from the rGO edge is attributed to a stable and unique aggregation of oxygen groups in the form of cyclic edge ethers. Such closely spaced electron beams from rGO offer prospects for novel applications and understanding the physics of linear electron sources.

**INTRODUCTION**

Electrons from a solid can be emitted into the vacuum by thermal excitation (i.e thermionic emission) [1], decreasing the work function [2-5], or applying sufficiently large electric fields (i.e. field emission) [1,6]. Most electron emitters such as x-ray sources utilize thermionic emission to generate high current electron beams but are problematic because they are unstable, bulky and cause heating of the surrounding device housing. Therefore cold cathode emitters that remain at room temperature and provide high emission current density at low electric fields are desirable. High aspect ratio materials such as carbon nanotubes and metallic tips exhibit excellent electron emission characteristics due to local field enhancement at the tip which decreases the barrier width, allowing electrons to tunnel into vacuum at low electric fields. Although low threshold fields (~ 1V/μm) have been reported for carbon field emitters of various types, careful experiments on individual or well separated emission sources have revealed that the onset of field emission is related only to the field enhancement factor ($\beta = h/r$ where h is the height and r is the radius) and the work function ($\Phi \sim 5eV$ for $sp^2$ carbon) of the emitter [7-9]. In unique cases, however, low threshold field emission is possible in ultra smooth surfaces, due to intrinsically lower (or even negative) work functions. For example, hydrogenation of diamond leads to a negative electron affinity surface [3-4, 10-11] and low threshold field emission can be readily observed [12-14]. However low carrier concentration in the conduction band due to its large energy gap limits the overall emission current from diamond cathodes, precluding their implementation into applications such as x-ray tubes [15-16], microwave amplifiers [17], and rocket thrusters [18-20] where high current density metallic cathodes that emit multiple closely packed electron beams at low threshold fields are required. In addition to low threshold fields and high current densities, emission of coherent electron beams is also technologically desirable. In rare instances, when the emission sites are located within the phase coherence length of the cathode material, as in hemispherical cap of a multi-walled carbon nanotube [21-22] or lithographically defined Pt tips that are located within ≤ 1nm of each other, coherent electron beams have been observed [23-24].

Here we demonstrate low threshold field emission from individual atomically thin reduced graphene oxide (rGO) sheets and show that it arises from a decrease in work function of the *edge* due to the creation of cyclic C-O-C ether groups in combination with geometric field enhancement. Furthermore the emission sites from rGO are densely packed, located only a few nanometers apart. Our investigation of the spatial distribution of emission sites with field ion microscopy (FIM) in combination with simulations of the emission patterns obtained using field emission microscopy (FEM) suggests that the emitted electron beams from several sites along the rGO edge interfere.

**RESULTS AND DISCUSSION**

Most of the reported field emission studies from graphene or rGO based cathodes have included a large number of flakes of varying dimensions and orientations, obscuring the intrinsic emission characteristics and mechanisms of individual graphene and rGO sheets by averaging effects [25-30]. In this study, we have fabricated field emitters consisting of individual rGO sheets prepared from suspensions of chemically derived graphene oxide (GO) [31-32], using a simple and reproducible methodology that results in individual vertically aligned rGO sheets (see Supplementary Information). The flakes were mounted on copper supports for the field emission experiments, as shown in Figure 1a. After cathode preparation, the flakes were annealed with repeated current induced heating prior to field emission measurements, thus thermally reducing the GO. This in-situ Joule heating or "pre-conditioning" step in field emission measurements is well known to heat carbon samples above ~1150 K, removing any adsorbed species that are present on the sample [8].

Field emission experiments were conducted by applying a bias voltage between the rGO cathode and the imaging anode plate, which was positioned perpendicular to the cathode plane to ensure that the electron emission occurred from the rGO edge (Figure 1b). The average threshold field required to emit currents of 1 nA was found to be less than 0.1 V/μm (Figure 1c). The field emission current was found to increase exponentially with the applied voltage, following the typical Fowler-Nordheim tunneling behavior [33] (Figure 1d, See Supplementary Information for details). The emission patterns obtained with FEM show that the emission sites are distributed over the entire rGO sheet edge.

Information about interactions between emitted electron beams was obtained from the FEM luminescence patterns on the anode above a critical applied voltage (~ 1 kV) between the anode and the cathode. FEM images obtained at different applied voltages are shown in Figure 2a. The increase in the number and brightness of the luminescent spots with applied voltage corresponds to the increase in the number of emission sites and current density. Interestingly, rGO cathodes exhibit emission patterns consisting of alternating bright and dark fringe bands as most readily observed in Figure 2a (iii) [the major bands are indicated by arrows], in contrast to conventional metal tip cathodes, which typically show a single to few bright circular spots [34-35]. Such fringe patterns with well defined bright and dim bands that remain relatively stable over time were only observed in single layer rGO cathodes, strongly suggesting that are not simply related to intensity fluctuations among emission sites but are a consequence of interactions among emitted beams from the rGO edge. Similar but much fewer fringes have been previously observed from multi-walled carbon nanotubes (MWNTs) [21-22, 36-37] and nano-split Pt emitters [23-24]. For comparison, we have also measured samples with multiple graphene layers (N > 10). The main features of the FEM patterns observed for multilayered samples consisted of multiple circular or elliptical spots (Figure 2b). Fower-Nordheim (FN) analysis of the field emission results reveals significantly larger geometric field enhancement factors for the single sheet rGO relative to the multilayered samples, suggesting that the difference in the FEM patterns is a direct consequence of the emitter being atomically thin (See Supplementary Information for details).

The electron emission site distribution along the atomically thin rGO edge was investigated with field ion microscopy (FIM) [1, 6] in which helium gas is physisorbed on to the cathode surface at liquid He temperatures and large negative bias is applied to the anode. Along the rGO edge where emission is most likely to occur, He gas molecules are ionized due to high field concentration and accelerated towards the phosphor screen, causing luminescence upon impact. The FIM image of emission sites from the rGO edge (indicated by the yellow line) consisting of an array of bright spots corresponding to emission sites responsible for the FEM patterns shown in Figure 2a (iii) is shown in Figure 2c. Analyses of the emitted electron beam angles reveal that the emission spots are aligned almost perpendicular to the corresponding fringe bands in the FEM patterns, suggesting that the emission occurs from the top of the rGO edge. Most importantly, the distance between the emission sites can be estimated to be on the order of a few nanometers based on the number and length of emission sites projected on the screen, the degree of defocusing of an electron beam, and separation distance between anode and cathode. This value is consistent with experimental electron microscopy measurements of dimensions of emission centers as well as FEM pattern simulations, as discussed below. The brightness profile of the emission spots along the yellow line in Figure 2c is plotted in Figure 2d. The profile clearly shows that the brightness peaks in Figure 2d occur at reasonably regularly spaced intervals, suggesting that emission sites are uniformly distributed along the rGO edge.

The FIM results indicate that the observed FEM fringe patterns may be due to interference among several electron beams originating from the rGO edge. The fringe patterns consist of a series of major and minor bands [figure 2a (iii)] unlike the typical circular or elliptical patterns observed for metallic tips or nanotube thin film emitters. To gain insight into electron beam interactions that give rise to the fringe bands in FEM patterns of rGO, we have simulated the projected intensity of multiple electron beams originating from an array of nanoscopic emission sources along an atomically thin edge (see Supplementary Information for details). The calculations used here are comparable to those used by Oshima *et al.*[21] to reproduce FEM patterns from MWNTs. In the simulations the size and spacing of emission sites along the edge can be adjusted. The best results in terms of being able to reproduce our experimentally observed FEM patterns were found for emission spots of ~ 1 nm in diameter separated by ~ 2 nm. Furthermore, we have found from our calculations that if the electron beams in such close proximity are coherent, then they readily interfere and that such interference leads to fringe patterns that are similar to those observed experimentally in Figure 2a (iii). Typically, interference between electron beams occurs only when the distance between emission sites is equal to or less than the coherence length of the material. For rGO, this length estimated from our transport studies is on the order of a few nanometers [38].

To verify the simulations, we compare the experimental FEM results with the calculated patterns. The intensity profile of the fringe pattern in Figure 2a (iii) plotted in Figure 2a (iv) shows that it consists of major and minor bands. In the simulations we found that for n = 2 emission sites, the FEM pattern consists of only major bright bands and no minor bands (see Supplementary Information). However for n > 2 emission sites, additional equally spaced minor bands appear between the major bands as observed in the calculated pattern shown in Figure 2a (v). Our calculations suggest that the number of emission sites on the rGO edge can be estimated from the number of minor bands observed in FEM patterns. In Figure 2a (iv) one minor band between the major bands can be seen, (arrows in 2a (iv) point to minor bands), indicating emission from three sites. Additional analysis of the intensity profiles of the FEM fringe patterns provided in the Supplementary Information shows interference from ≥ 4 electron beams. It can be further noticed that the intensity, separation, and the width of the bands remain nearly constant in our experimental FEM measurements, making it highly unlikely that anomalous effects such as intensity distribution among emission sites could be responsible. This suggests that more than two electron beams are constantly interfering and several emission sites are responsible for the observed patterns (See Supplementary Information for details).

The results presented thus far reveal that the low threshold field emission occurs from multiple sites on the rGO edge. Furthermore the FEM and FIM results indicate that the closely spaced electron beams interfere to produce fringe patterns comparable to those from MWNTs [21]. To determine the structure of the edges, we have performed extensive TEM imaging analyses. Imaging atomically thin edges in TEM is challenging because the time required to locate and focus on the edge leads to high electron dosage exposure, causing it to evolve and change. Nevertheless, we have been able to image edge structures sufficiently quickly before significant evaporation or structural alteration could occur. A low magnification image of a single layer edge is shown in the Supplementary Information Figure S6.

A high resolution TEM image of the edge perturbed by a kink makes it possible to investigate its termination, as shown in Figure 3a. Based on the contrast difference between carbon and oxygen atoms in the TEM, we can deduce that the structure of the edge shown in Figure 3a consists of C-O-C cyclic ethers. An edge consisting of purely carbon atoms is shown in Figure 3b. The contrast intensity profile (without correcting for electron beam probe contributions) in Figure 3c shows a clear difference between the oxygen and carbon atoms with peaks corresponding to positions of oxygen atoms along the line in Figure 3a. In Figure 3d, such a contrast difference is not observed in an edge containing primarily carbon atoms. Using this information, the oxygen atoms are indicated along the edge in Figure 3e by red dots and a model of the structure shown in Figure 3f. These TEM observations are in excellent agreement with our recent IR spectroscopic investigation on similarly prepared GO [39], which clearly revealed that upon thermal annealing, the oxygen functional groups form C-O-C cyclic ether groups at the sheet edges. In that work, the observation of a giant IR absorption peak at 800 cm$^{-1}$ was critical to establish cyclic ether termination of atomically straight edges. The unusual absorption at 800 cm$^{-1}$ was shown by first principles calculations to arise from density of mobile states created by the asymmetric stretching mode of the C-O-C groups at the rGO edge (See Ref 39). Additionally, the activation of such C-O-C metallic states requires adjacent oxygen atoms to be greater than ~ 7 in the chain and accompanied by pristine sp$^2$ graphene region near the edge. The IR spectroscopy and simulations quantification results a (see Supplementary Information for detailed analysis of the C-O-C concentration along the rGO edge) also indicate that only a small fraction (~ 1%) of the sheet edges are straight and decorated with ether groups, making their observations by TEM particularly challenging as noted above.

To determine how the C-O-C ether groups at the rGO edge influence the field emission properties, we have performed density functional theory (DFT) calculations to determine the work function of the rGO edge as a function of the oxygen content. The results of the DFT calculations are shown in Figures 3g and 3h, which show the variation of the work function with distance from within the rGO sheet to the edge for varying applied electric fields. The case for pure graphene is given in Figure 3g, which shows that in the absence of an applied electric field the work function at the edge is 5.45eV, significantly larger than that of graphite. The work function decreases with applied electric field but even at extremely large fields (1 V/Å = $10^4$ V/μm), the barrier remains high at 4.2 eV. In contrast, the work function of the rGO edge containing C-O-C functional groups with ≥ 7 adjacent oxygen atoms is substantially lower at 3.70eV even in the absence of an applied field. Thus, the presence of C-O-C groups at the rGO edge significantly reduces the barrier for electron emission into vacuum compared to a pristine graphene edge. In addition to the two cases shown in Figures 3g and 3h, we have also monitored the barrier heights as a function of the oxygen atoms along the edge and found that a minimum of ~ 7 oxygen atoms are required to lower the work function (see Supplementary Information). This suggests that the exceptionally low threshold field emission observed in this study is closely related to the presence of metallic density of states in low work function graphene structures with straight edge cyclic ether termination (See Ref 39 for detailed description of the density of states and electronic structure calculations of C-O-C chains on rGO edge).

We surmise that field emission is uniform along the rGO edge and occurs through sites that are localized at low work function C-O-C ether chain regions, as summarized in the self-consistent model illustrated in Figure 4. The low work function C-O-C chain dimensions are ~ 1 nm, consistent with the size of the emission sites used in the FEM simulations. Assuming that the field emission occurs from the low function regions on the edge of rGO and the fact that the electronic structures of C-O-C ether chains along the rGO edge are very similar, it is not surprising that the emitted electron beams from such sites would have similar phase characteristics. This combined with the fact that the FEM simulations and FIM results suggest that the maximum separation between the emission sites is within a few nanometers (within the coherence length of rGO from Ref 38) satisfies the basic conditions for the interference of the emitted electron beams.

**CONCLUSIONS**

We have demonstrated exceptionally low threshold field emission from atomically thin edges of rGO. The edges are found to provide an array of emission sites in the form of low work function C-O-C ether chains from which multiple electron beams are simultaneously emitted. The emission from rGO edges occurs from localized C-O-C chains with similar electronic structures so that the emitted electron beams appear to be coherent as indicated by the fact that they interfere with each other and produce fringe patterns in field emission microscopy. The enhanced field emission characteristics are attributed to the termination of straight edges by cyclic ether that constitute the most stable form of oxygen in rGO. These results demonstrate that atomically thin edges are excellent linear sources of high-density electron beams. Furthermore, coherent electron beams from the rGO may open up new and interesting physics as well as novel applications such as nano-lithography, vacuum electronics, and electron optics devices.

**METHODS**

**Sample preparation.** Graphene was prepared chemically via modified Hummer's method. The graphite powders (Branwell Graphite Inc.) with size of > 425 μm were chemically oxidized, exfoliated, and purified by repeated centrifugation. Aqueous chemically derived graphene solution was then drop cast onto Cu grids. Parallel grids instead of mesh grids were used to exclude possibility of film deposition on the side of the bars. Prior to casting, the grid lines were cut from the outer ring using focused ion beam (FIB) (DB Strata 235 Dual-Beam FIB). When the solution dried, suspended GO film formed over the prepatterned gap. Suspended GO flakes were reduced via hydrazine monohydrate vapor and thermal annealing at 200 $^{o}$C in vacuum for 5 and 20 hours, respectively. The individual grid line was carefully removed obtain a bar with graphene film on its end. The bar was fixed onto the Tungsten (W) needle sample stage using silver paint and inserted into the ultrahigh vacuum chamber for measurements.

**FEM and FIM measurements.** The base pressure of the measurement system was ~1 x $10^{-7}$ Pa. Phosphor-coated indium-tin oxide (ITO)/glass plate was placed about 22 mm from the sample. FEM and FIM were performed on the same area of an identical sample without breaking vacuum. This minimizes adsorptions on the sample which would prevent direct comparison between the two techniques. A few cycles of voltage ramps up to 4 kV were applied prior to FEM imaging until stable I-V characteristics with emission currents in the range of $10^{-9}$-$10^{-5}$ A were obtained. The emission patterns on the phosphor screen were captured by a CMOS camera. Measurements were performed both at room temperature and at ~7 K.

For FIM, Micro Channel plate Photomultiplier (MCP) was placed in front of the phosphor screen between the anode and the sample to enhance the image visibility. The sample was cooled down to ~7 K, and He gas was introduced into the chamber to allow its condensation onto the sample surface. A negative bias of ~10 kV was applied to ITO anode relative to the grounded graphene to induce field ionization of the He. The pressure during the FIM measurement was ~$10^{-3}$ Pa. The luminescence of the phosphor was captured by a CMOS camera.

**ACKNOWLEDGEMENTS**

The authors would like to acknowledge Dr. T. Yamada of National Institute of Advanced Industrial Science and Technology, Japan for discussion on the analysis of field emission characteristics. The authors would also like to thank Prof. E. Saiz of Imperial College London for discussions on the interpretation of the interference patterns. We also acknowledge Prof. D. Natelson of Rice University, Dr. H. Najafov of Rutgers University for discussions. This work was funded by the National Science Foundation CAREER Award (ECS 0543867), Grant-in-Aid for Scientific Research A (No.19206008) & Grant-in-Aid for Young Scientists B (No.20410108) from the Ministry of Education, Culture, Sports, Science and Technology of Japan. We also acknowledge the financial support from Center for Advanced Structural Ceramics (CASC) at Imperial College London. GE would like to acknowledge the Royal Society for the Newton International Fellowship. MC acknowledges support from the Royal Society through the Wolfson Merit Award. YC acknowledges NSF (CHE-0911197).

**FIGURE CAPTIONS**

**Fig. 1.** Experimental set up of the field emission measurements. **(a)** Typical SEM image of rGO sheet mounted on a copper support bar. **(b)** Schematic of the FEM measurement set up. Voltages up to 4 kV were applied between the imaging anode and the cathode for FEM. For FIM, opposite bias is applied and helium ions (instead of electrons) produce the image on the screen. **(c)** *I-V* field emission characteristics and corresponding **(d)** Fowler-Nordheinm plot for rGO (red) and the multilayered (black) samples.

**Fig. 2.** FEM and FIM images of atomically thin and the multilayered rGO sheets at different applied voltages. Scale bars indicate 5 mm at the imaging anode. **(a)** FEM patterns as a function of applied voltage. Fringe patterns were observed at anode voltage of $\geq$ 2kV. The arrows in (iii) indicate the major bands in the FEM pattern. (iv)

Intensity profile of (iii) across the fringe region. The arrows indicate the position of minor bands. (v) Simulated FEM patterns for the case of three aligned emission sites. **(b)** FEM patterns for a multilayered graphitic cathode as a function of the applied voltage showing elliptical and circular FEM patterns and absence of fringes. **(c)** FIM image showing emission sites on the rGO edge (measurements taken at 40 K). Numerous emission spots (>10) along the line corresponding to the rGO sheet edge can be seen. **(d)** Brightness profile of the emission sites along the yellow line in (c).

**Fig. 3.** TEM images and work function of rGO edges. Representative atomic resolution TEM images of an edge **(a)** containing oxygen functional groups in the form of C-O-C éther chains and **(b)** an edge containing only carbon atoms. **(c)** Contrast intensity profile along the line in (a) showing peaks corresponding the oxygen atoms. **(d)** Contrast intensity profile along the line shown in (b) showing absence of peaks, indicating that most of the atoms are carbon. **(e)** TEM image with red dots indicating the positions of the oxygen atoms. **(f)** Atomic structure of the edge illustrated in ball and stick model. Red and gray balls indicate oxygen and carbon atoms, respectively. Density functional calculations showing the variation of the work function from within the sheet to its edge at different applied field for **(g)** pure graphene and **(h)** rGO.

**Fig. 4**. Schematic model of the field emission from rGO edge. Field emission occurs from low work function C-O-C functionalized regions containing at least 7 oxygen atoms as indicated by the red rectangles. The prevalent non-oxygenated edges are higher work function. Interference from sufficiently close electron beams is also depicted.

**REFERENCES**

1. A. Modinos. Field, Thermionic and Secondary Electron Emission Spectroscopy. (Plenum Press, New York, 1984).
2. R.L. Bell. Negative Electron Affinity Devices. (Clarendon, Oxford, 1973).
3. F.J. Himpsel, J.A. Knapp, J.A. VanVechten and D.E. Eastman. Quantum photoyield of diamond(111)-A stable negative-affinity emitter. *Phys. Rev. B* 20, 621-627 (1979).
4. B.B. Pate, M.H. Hecht, C. Binns, I. Lindau and W.E. Spicer. Photoemission and photon-stimulated ion desorption studies of diamond(111): Hydrogen. *J. Vac. Sci. Technol.* 21, 364-367 (1982).
5. E. Eyckeler, W. Monch, T.U. Kampen, R. Dimitrov, O. Ambacher and M. Stutzmann. Negative electron affinity of cesiated p-GaN (0001) surfaces. *J. Vac. Sci. Technol. B* 16, 2224-2228 (1998).
6. R. Gomer. Field Emission and Field Ionization (Harvard University Press, Cambridge, Mass., 1961).
7. K.B.K. Teo, M. Chhowalla, G.A.J. Amaratunga, W.I. Milne, P. Legagneux, G. Pirio, L. Gangloff, D. Pribat, V. Semet, B. Vu Thien, W.H. Bruenger, J. Eichholz, H. Hanssen, D. Friedrich, S.B. Lee, D.G. Hasko and H. Ahmed. Fabrication and electrical characteristics of carbon nanotube-based microcathodes for use in a parallel electron-beam lithography system. *J. Vac. Sci. Technol. B* 21, 693-697 (2003).
8. S.T. Purcell, P. Vincent, C. Journet and V.T. Binh. Hot nanotubes: Stable heating of individual multiwall carbon nanotubes to 2000 K induced by the field-emission current. *Phys. Rev. Lett.* 88, 105502 (2002).
9. J.-M. Bonard, K.A. Dean, B.F. Coll, C. Klinke. Field Emission of Individual Carbon Nanotube in the Scanning Electron Microscope. *Phys. Rev. Lett.* 89, 197602 (2002).
10. J. van der Weide, Z. Zhang, P.K. Baumann, M.G. Wensell, J. Bernholc, R.J. Nemanich. Negative-electron-affinity effects on the diamond (100) surface. *Phys. Rev. B* 50, 5803–5806 (1994).
11. J.B. Cui, J. Ristein and L. Ley. Low-threshold electron emission from diamond. *Phys. Rev. B* 60, 16135–16142 (1999).
12. K. Okano, S. Koizumi, S.R.P. Silva and G.A.J. Amaratunga. Low-threshold cold cathodes made of nitrogen-doped chemical-vapor-deposited diamond. *Nature* 381, 140-141 (1996).
13. M.W. Geis, N.N. Efremow, K.E. Krohn, J.C. Twichell, T.M. Lyszczarz, R. Kalish, J.A. Greer and M.D. Tabat. A new surface electron-emission mechanism in diamond cathodes. *Nature* 393, 431-435 (1998).
14. W. Zhu, G.P. Kochanski and S. Jin. Low-field electron emission from undoped nanostructured diamond. *Science* 282, 1471-1473 (1998).

15. H. Sugie, M. Tanemura, V. Fillip, K. Iwata, K. Takahashi and F. Okuyama. Carbon nanotubes as electron source in an x-ray tube. *Appl. Phys. Lett.* 78, 3578-2580 (2001).
16. G.Z. Yue, Q. Qiu, Bo Gao, Y. Cheng, J. Zhang, H. Shimoda, S. Chang, J.P. Lu and O. Zhou. Generation of continuous and pulsed diagnostic imaging x-ray radiation using a carbon-nanotube-based field-emission cathode. *Appl. Phys. Lett.* 81, 355-357 (2002).
17. K.B.K. Teo, E. Minoux, L. Hudanski, F. Peauger, J.-P. Schnell, L. Gangloff, P. Legagneux, D. Dieumegard, G. A. J. Amaratunga and W. I. Milne. Carbon nanotube as cold cathodes. *Nature* 437, 968 (2005).
18. M. Gamero-Castano, V. Hruby, P. Falkos, D. Carnahan, B. Ondrusek and D. Lorent. Electron Field Emission from Carbon Nanotubes, and its Relevance in Space Applications. *36th AIAA/ASME/SAE/ASEC Joint Propulsion Conference and Exhibit* AIAA 2000-3263 (2000).
19. B. Gassend, L.F. Velasquez-Garcia, A.I. Akinwande and M. Martinez-Sanchez. Mechanical Assembly of Electrospray Thruster Grid. *Proc. of the 29th International Electric Propulsion Conference* IEPC 2005-101 (2005).
20. L.F. Velasquez-Garcia, A.I. Akinwande and M. Martinez-Sanchez. A Planar Array of Micro-fabricated Electrospray Emitters for Thruster Applications. *J. of MicroElectro-Mechanical Systems* 15, 1272-1280 (2006).
21. C. Oshima, K. Mastuda, T. Kona, Y. Mogami, M. Komaki, Y. Murata, T. Yamashita, T. Kuzumaki and Y. Horiike. Young's interference of electrons in field emission patterns. *Phys. Rev. Lett.* 88, 038301 (2002).
22. T. Kuzumaki, Y. Horiike, T. Kizuka, T. Kona, C. Oshima and Y. Mitsuda. The dynamic observation of the field emission site of electrons on a carbon nanotube tip. *Diam. Rel. Mater.* 13, 1907-1913 (2004).
23. K. Murakami, F. Wakaya and M. Takai. Observation of fringelike electron-emission pattern in field emission from Pt field emitter fabricated by electron-beam-induced deposition. *J. Vac. Sci. Technol. B* 25, 1310-1314 (2007).
24. K. Murakami, T. Matsuo, F. Wakaya and M. Takai. Electron-wave interference induced by electrons emitted from Pt field emitter fabricated by focused-ion-beam-induced deposition. *J. Vac. Sci. Technol. B* 28, C2A9 (2010).
25. S.G. Wang, J.J. Wang, P. Miraldo, M.Y. Zhu, R. Outlaw, K. Hou, X. Zhao, B.C. Holloway, D. Manos, T. Tyler, O. Shenderova, M. Ray, J. Dalton and G. McGuire. High field emission reproducibility and stability of carbon nanosheets and nanosheet-based backgated triode emission devices. *Appl. Phys. Lett.* 89, 183103 (2006).
26. S. Watcharotone, R.S. Ruoff and F.H.Read. Possibilities for Graphene for Field Emission: Modeling Studies using the BEM. *Phys. Proc.* 1, 71-75 (2008).
27. G. Eda, H.E. Unalan, N. Rupesinghe, G. Amaratunga and M. Chhowalla. Field emission from graphene based composite thin films. *Appl. Phys. Lett.* 93, 233502 (2008).

28. M. Alexander, K. Raymond, V. Annick, C. Manish Pal, V. Alexander and H. Chris Van. Field emission from vertically aligned few-layer graphene. *J. Appl. Phys.* 104, 084301 (2008).
29. Z.-S. Wu, S. Pei, W. Ren, D. Tang, L. Gao, B. Liu, F. Li, C. Liu and H.-M. Cheng. Field emission of single-layer graphene films prepared by electrophoretic deposition. *Adv. Mater.* 21, 1756-1760 (2009).
30. Z. Xiao, J. She, S. Deng, Z. Tang, Z. Li, J. Lu and N. Xu. Field Electron Emission Characteristics and Physical Mechanism of Individual Single-Layer Graphene. *ACS Nano*, 4, 6332–633 (2010).
31. G. Eda, G. Fanchini and M. Chhowalla. Large-area ultrathin films of reduced graphene oxide as a transparent and flexible electronic material. *Nature Nanotech.* 3, 270-274 (2008).
32. S. Stankovich, D.A. Dikin, R.D. Piner, K.A. Kohlhaas, A. Kleinhammes, Y. Jia, Y. Wu, S.T. Nguyen and R.S. Ruoff. Synthesis of graphene-based nanosheets via chemical reduction of exfoliated graphite oxide. *Carbon* 45, 1558-1565 (2007).
33. R. H. Fowler and L. Nordheim. Electron emission in intense fields. *Proc. R. Soc. Lond. A* 119, 173-181 (1928).
34. E. Rokuta, T. Itagaki, T. Ishikawa, B.L. Cho, H.S. Kuo, T.T. Tsong and C. Oshima. Single-atom coherent field electron emitters for practical application to electron microscopy: Buildup controllability, self-repairing function and demountable characteristic. *Appl. Surf. Sci.* 252, 3686-3691 (2006).
35. K. Murakami, N. Yamasaki, S. Abo, F. Wakaya and M. Takai. Observation of electron emission pattern from nanosplit emitter fabricated using beam assisted process. *J. Vac. Sci. Technol. B* 23, 735-740 (2005).
36. Y. Saito, K. Hata and T. Murata. Field emission patterns originating from pentagons at the tip of a carbon nanotube. *Jpn. J. Appl. Phys.* 39, L271-272 (2000).
37. H. Tanaka, S. Akita, L. Pan and Y. Nakayama. Daisylike Field-Emission Images from Standalone Open-Ended Carbon Nanotube. *Jpn. J. Appl. Phys.* 43, L197-199 (2004).
38. G.Eda, C.Mattevi, H.Yamaguchi, H.Kim and M.Chhowalla. Insulator to Semimetal Transition in Graphene Oxide. *J. Phys. Chem. C* 113, 15768-15771 (2009).
39. M. Acik, G. Lee, C. Mattevi, M. Chhowalla, K. Cho and Y.J. Chabal. Unusual infrared-absorption mechanism in thermally reduced graphene oxide. *Nature Materials* 9, 840–845, (2010).

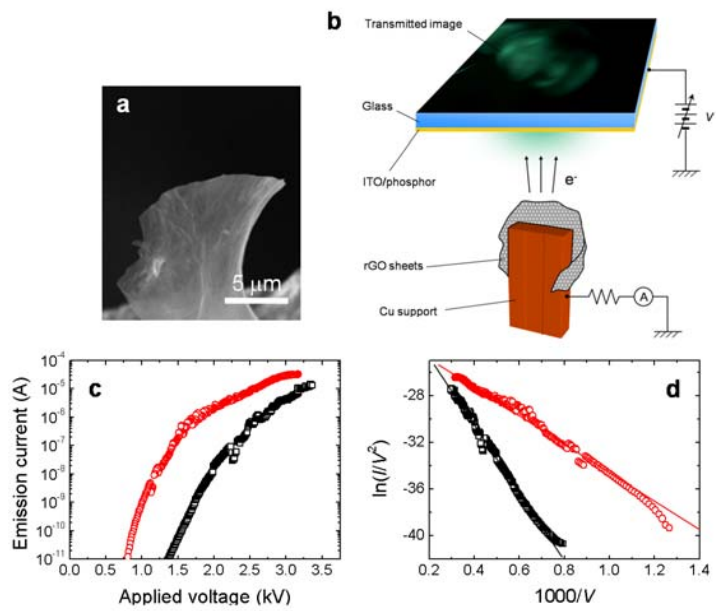

Figure 1  H.Yamaguchi *et al.*

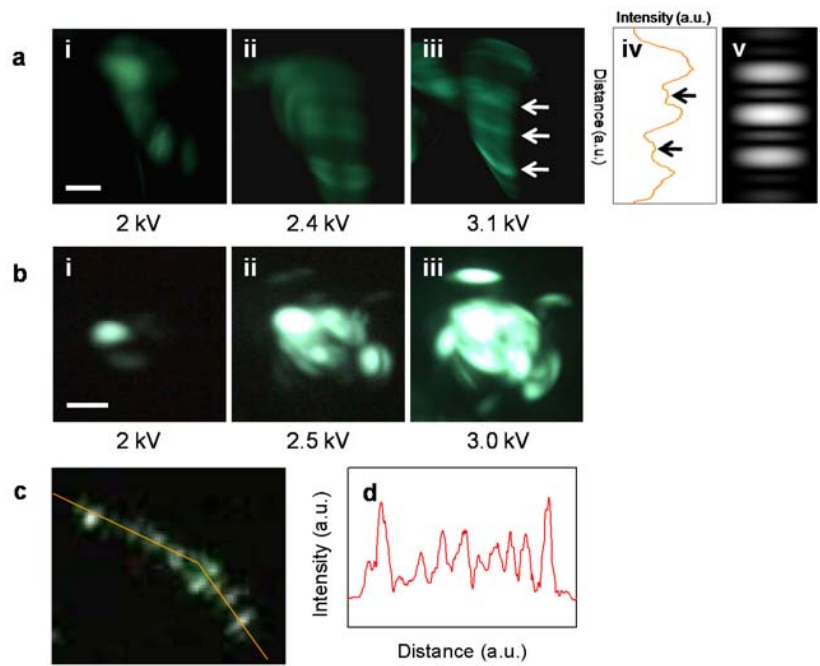

Figure 2 H.Yamaguchi *et al*.

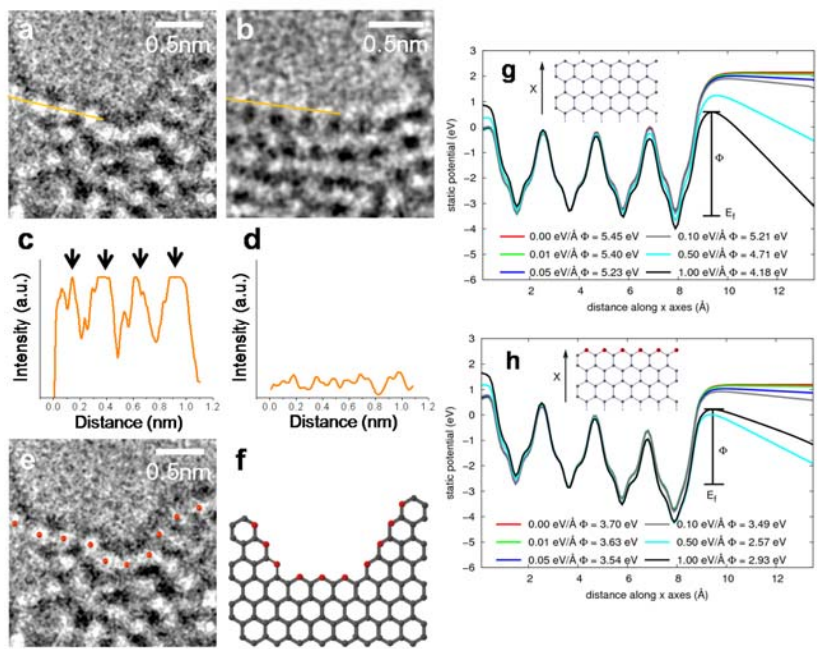

Figure 3  H.Yamaguchi *et al.*

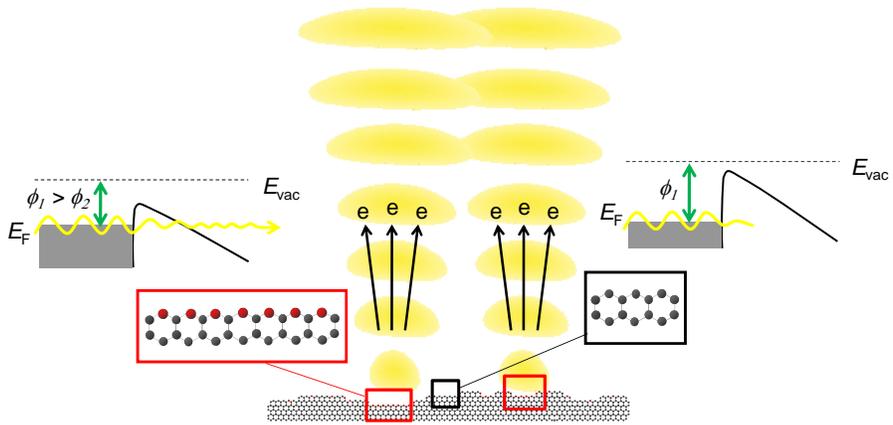

Figure 4  H.Yamaguchi *et al.*